\documentclass[12pt,preprint]{aastex}

\shorttitle{The Amplitude of Mass Fluctuations}
\shortauthors{Bahcall \& Bode}

\received{2002 December 16}

\begin{document}

\title{ The Amplitude of Mass Fluctuations }

\author{ Neta A. Bahcall and Paul Bode }
\affil{Princeton University Observatory, Princeton, NJ 08544-1001}
\email{neta@astro.princeton.edu, bode@astro.princeton.edu}

\begin{abstract}
We determine the linear amplitude of mass fluctuations in the universe, 
$\sigma_8$, from the abundance of massive clusters at redshifts
$z=0.5-0.8$. 
The evolution of massive clusters depends exponentially on the amplitude 
of mass fluctuations and thus provides a powerful measure of this important 
cosmological parameter. The relatively high abundance of massive clusters 
observed at $z>0.5$, and the relatively slow evolution of their 
abundance with time, suggest a high amplitude of mass fluctuations:
$\sigma_8 =0.9\pm 10$\% for $\Omega_m =0.4$, increasing slightly to
$\sigma_8 =0.95$ for $\Omega_m =0.25$ and
$\sigma_8 =1.0$ for $\Omega_m =0.1$ (flat CDM models).
We use the cluster abundance observed at $z=0.5-0.8$ to derive 
a normalization relation from the high-redshift clusters, which is only weakly 
dependent on $\Omega_m$: $\sigma_8 \Omega_m^{0.14}=0.78\pm 0.08$.
When combined with recent constraints
from the present-day cluster mass function, 
$\sigma_8 \Omega_m^{0.6}=0.33\pm 0.03$, we find $\sigma_8 =0.98\pm 0.1$ and
$\Omega_m =0.17\pm 0.05$. Low $\sigma_8$ values ($\la 0.7$) 
are unlikely; 
they produce an order of magnitude fewer massive clusters than observed.
\end{abstract}

\keywords{cosmological parameters --- galaxies: clusters: general}

\section{Introduction} \label{sec:intro}

The amplitude of mass fluctuations is a fundamental cosmological parameter 
that describes the normalization of the linear spectrum of mass
fluctuations 
in the early universe -- the spectrum that seeded galaxy formation. 
The abundance of massive clusters depends exponentially on this parameter 
(assuming Gaussian initial fluctuations), because
a high amplitude of mass fluctuations 
forms structure rapidly at early times, while a lower amplitude forms 
structure more slowly.  The most massive systems 
($\sim 10^{15}h^{-1}M_\odot$), which
take the longest time to form and grow,
did not exist at early times
if the initial amplitude of mass fluctuations is low,
but rather formed only recently.

The amplitude parameter, denoted $\sigma_8$ when 
referring to the {\it rms} linear density fluctuation 
in spheres of radius $8h^{-1}$Mpc at $z=0$,
is not easily determined since the mass distribution 
cannot be directly 
observed. As a result, this parameter is not yet accurately known.
Recent observations suggest an amplitude 
that ranges in value from  $\sigma_8\sim 0.7$ to a 
`high' value of $\sigma_8\sim 0.9-1$. 
While the difference in the reported values is only around
50\%, the impact on 
structure formation and evolution is much larger, since the latter depends 
exponentially on $\sigma_8^2$. The low amplitude values are suggested by 
current observations of the CMB spectrum of fluctuations 
\citep{netterfield, sievers02, bondea02, ruhlea02}.
However, this 
$\sigma_8$ determination is degenerate with the unknown optical depth at
reionization: 
if the optical depth were underestimated, then
$\sigma_8$ would be as well\footnote{See note at end of paper}.  
Recent observations of the 
present-day cluster abundance as well as cosmic shear lensing measurements 
have also suggested that $\sigma_8\sim 0.7$
\citep[e.g.][]{jarvis03, hamana02, sel01}.
However, these measures provide a degenerate relation 
between the amplitude $\sigma_8$ and the mass-density parameter 
$\Omega_m$: $\sigma_8 \Omega_m^{0.6}\approx 0.33$ 
\citep{ike02, SDSSmf03, jarvis03, sel01}.
The amplitude
$\sigma_8\sim 0.7$ is implied only if 
$\Omega_m \sim 0.3$. If
$\Omega_m \sim 0.2$, as is suggested by some observations 
\citep[e.g.][]{CYE97, bah98, bah00, wklc01, ike02, rei02},
then the amplitude is $\sigma_8\sim 0.9-1$.  Early results 
from the Sloan Digital Sky Survey (SDSS) cluster data \citep{SDSSmf03}
use the shape of the observed cluster mass function to
break the degeneracy between the parameters and find 
$\sigma_8 =0.9^{+0.3}_{-0.2}$ and $\Omega_m =0.19^{+0.08}_{-0.07}$.
Similar results have recently
been obtained from the temperature function of a large sample of
X-ray clusters \citep{ike02, rei02}.
Most of the recent cluster normalization observations,
as well as cosmic shear lensing measurements 
suggest $\sigma_8\simeq 0.9-1$ if $\Omega_m \simeq 0.2$
\citep[][and the references above]{jarvis03, hamana02}.
Combining current CMB measurements with the SDSS cluster
mass function yields intermediate values of
$\sigma_8 =0.76\pm 0.09$ and $\Omega_m =0.26^{+0.06}_{-0.07}$
\citep{MBBS03}.

The evolution of cluster abundance with time, especially for the most massive
clusters, breaks the degeneracy between $\sigma_8$ and $\Omega_m$
\citep[e.g.][]{PDJ89, ECF96, OB97, BFC97, CMYE97, bah98, DV99, Henry00}.
This evolution depends 
strongly on $\sigma_8$,
and only weakly on $\Omega_m$ or other parameters.
The expected abundance of massive clusters 
at $z\sim 0.5-1$ differs between Gaussian models with $\sigma_8=0.6$
and $\sigma_8=1$
by orders-of-magnitude, nearly independently of other parameters
\citep[e.g.]{FBC97}.
Therefore, this method provides a uniquely
powerful tool in estimating the amplitude $\sigma_8$.

In this paper we use the abundance of the most massive clusters 
observed at $z\sim 0.5-0.8$  to place a strong limit on $\sigma_8$.

\section{The Evolution of Cluster Abundance} \label{sec:evol}

A flat $\Omega_m=0.3$ cold-dark-matter 
universe (LCDM) with $\sigma_8=0.6$ predicts $\sim 10^3$
fewer massive clusters 
(with mass $\sim 10^{15}h^{-1}M_\odot$)
at $z\approx 0.8$ than a universe with
$\sigma_8=1$ \citep[e.g.][]{bah98, bod01}.
We use massive clusters observed at $z\approx 0.5$ 
and $z\approx 0.8$  
\citep[as compiled by][]{bah98}
to set limits on $\sigma_8$.
These three clusters, detected originally in X-rays by the EMSS survey 
\citep{Henryea92, LupGio95},
have masses larger than 
$8\times 10^{14}h^{-1}M_\odot$ 
within a radius of 1.5$h^{-1}$ comoving Mpc. The masses are 
determined from gravitational weak lensing observations 
(for two of the three clusters), as well as from the observed temperatures 
($T \gtrsim$8 Kev) and velocity dispersions ($\gtrsim$1200 Km/s) 
of the clusters. 
All clusters have a measured S-Z effect 
\citep{Grego01, carea01}.
These clusters have been
conservatively selected, with
mass measurements available from several independent methods, all yielding 
consistent results (see \citet{bah98} for details, including the 
consistency of the mass determinations from different methods, and 
the relevant abundances at $z\approx 0.6$ and 0.8). Since 
only a threshold cluster
mass is used in the analysis below (i.e., not individual cluster masses),
a conservative mass threshold of 
$8\times 10^{14}h^{-1}M_\odot$  is used;
the clusters are well above this threshold. 
The resulting abundances are
$n_{cl} =1.4^{+1.1}_{-0.9}\times 10^{-8}h^3$Mpc$^{-3}$
at $z=0.5-0.65$, 
and $1.4^{+1.4}_{-1.1}\times 10^{-8}h^3$Mpc$^{-3}$
at $z=0.65-0.9$. 
The error bars represent 68\% confidence level assuming Poisson 
statistics and equal likelihood for each log($n_{cl}$).
The volume searched and hence the number density depends on
the assumed cosmology;
these numbers are for an LCDM model with $\Omega_m = 0.3$.

For purposes of comparison,
the abundance of massive clusters at $z\approx 0$ 
and at $z\approx 0.38$ is obtained
from the temperature function observed by \citet{ike02}
(for $z\approx 0$) and \citet{Henry00}
(for $z\approx 0.38$). Here we convert 
our threshold mass of 
$8\times 10^{14}h^{-1}M_\odot$  (within 1.5$h^{-1}$ comoving Mpc) to 
temperature using the 
mean relation between cluster mass (observed directly from 
weak gravitational lensing) and temperature:
$M$($\leq 1h^{-1}$ Mpc) = 0.95$T$(kev) $10^{14}h^{-1} M_\odot$
(\citet{bahs02}; see also \citet{HOvK98, CYE97}).
These observations indicate that cluster masses determined from lensing
are consistent on average with those derived from X-ray temperatures,
with an {\it rms} scatter of $\sim$20\%.
For the small extrapolation 
to radius 1.5$h^{-1}$ Mpc comoving, we use the observed 
cluster profile on these scales: $M(<R) \sim R^{0.6}$
\citep{CYE97, fis97}.
This allows us to determine the observed cluster 
abundance at $z\simeq 0$ and $z\simeq 0.38$ for the relevant mass clusters. 
We find $n_{cl} =1.1^{+1.1}_{-0.7}\times 10^{-7}h^3$Mpc$^{-3}$
at $z=0.05$ \citep{ike02} and 
$n_{cl} =1.7^{+1.7}_{-1.1}\times 10^{-8}h^3$Mpc$^{-3}$
at $z=0.3-0.5$ \citep{Henry00};  the error bars allow for
statistical uncertainty as well as the uncertainty in the mass threshold.
These numbers assume an $\Omega_m = 0.3$ LCDM cosmology.

The observed abundance of these massive clusters as a function of 
redshift is presented in Figure \ref{fig:sig8plot}. 
In the next section, we compare the data with 
semianalytic predictions and constrain the 
allowed range of parameters. 
First, we use only the high redshift ($z>0.5$) 
cluster abundance (where mass thresholds are determined by 
multiple methods, as discussed above).
Then we take advantage
of the full evolution of the abundance, from $z\sim 0$ 
to $z\sim 0.8$, by combining 
the high-redshift result with the independent $z\approx 0$ cluster 
normalization relation obtained from other studies.
We discuss these analyses in the following section.

\section{Comparison with Predicted Densities} \label{sec:comp}

For a given cosmological model, the expected cluster mass
function can be predicted using recent improvements to the
Press-Schechter formalism.  We follow the procedure outlined in
the Appendix of \citet{HuKrav02}.  For a given choice of parameters,
the linear matter power spectrum is calculated with the publicly
available CMBFAST code \citep{CMBFAST96}.  This allows a prediction
of the mass function, using a fixed mean overdensity as the
definition of mass.  This mass can then be extrapolated to 
1.5 $h^{-1}$Mpc comoving,
assuming an NFW density profile.  The analytic predictions were 
compared to the N-body results of \citet{bod01}, and 
the resulting cluster abundances were found to
agree within 20\% out to redshift $z=1$.

\begin{figure}
\plotone{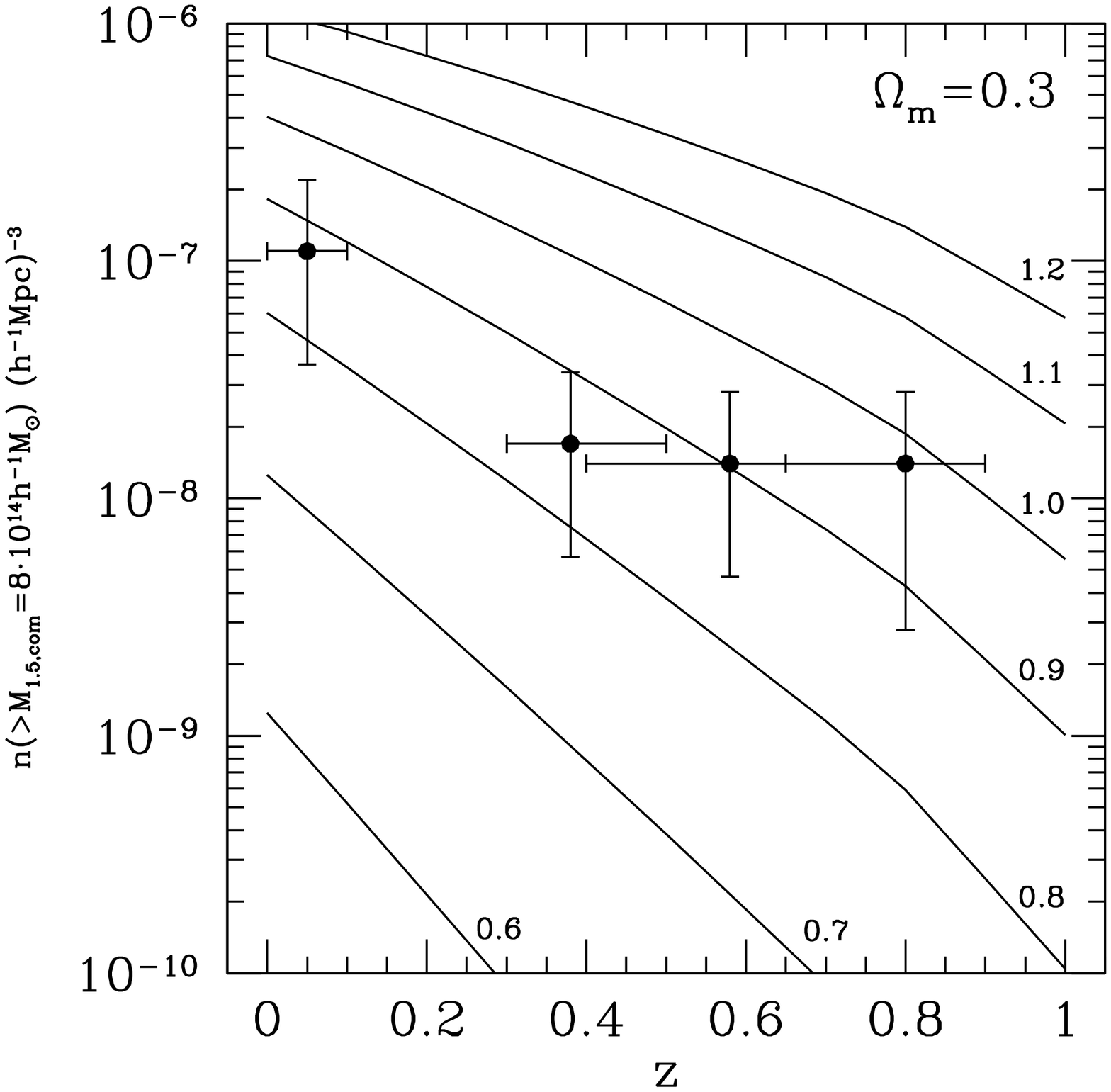}
\caption{Evolution of cluster abundance with redshift,
for clusters with mass 
$M_{1.5,com}\geq 8\times 10^{14} h^{-1}M_\odot$
(within a comoving radius of $1.5h^{-1}$Mpc).
Dots with error bars are the data
as described in the text.  The lines are the predicted number
density assuming a flat $\Omega_m=0.3$ cosmology;  each line
is labeled with the $\sigma_8$ used.
\label{fig:sig8plot}}
\end{figure}

The first result is presented in 
Figure \ref{fig:sig8plot}, where we compare the observed
cluster abundance 
as a function of redshift with that expected from LCDM models.
Here we use LCDM  with the `concordance' value of $\Omega_m =0.3$ 
\citep[][with $h=0.72$ and $n=1$]{bah99},
but with different 
amplitudes $\sigma_8$ ranging from 0.6 to 1.2 
(bottom to top curves).
Comparison of data and models shows that 
if $\Omega_m =0.3$, then $\sigma_8$ has 
to be relatively high, $\sigma_8 =0.9\pm 0.1$, 
in order to produce the observed
abundance of clusters at all
redshifts. Low normalization values of 
$\sigma_8\approx 0.7$--- which would be required for
$\Omega_m=0.3$ by
the present-day cluster abundance and the
cosmic shear lensing measurements
discussed in \S\ref{sec:evol}--- produce an order-of-magnitude 
too few massive clusters as
compared with observations at any redshift. 
Therefore, for $\Omega_m$=0.3, $\sigma_8$ needs to be $\gtrsim 0.9$.

\begin{figure}
\plotone{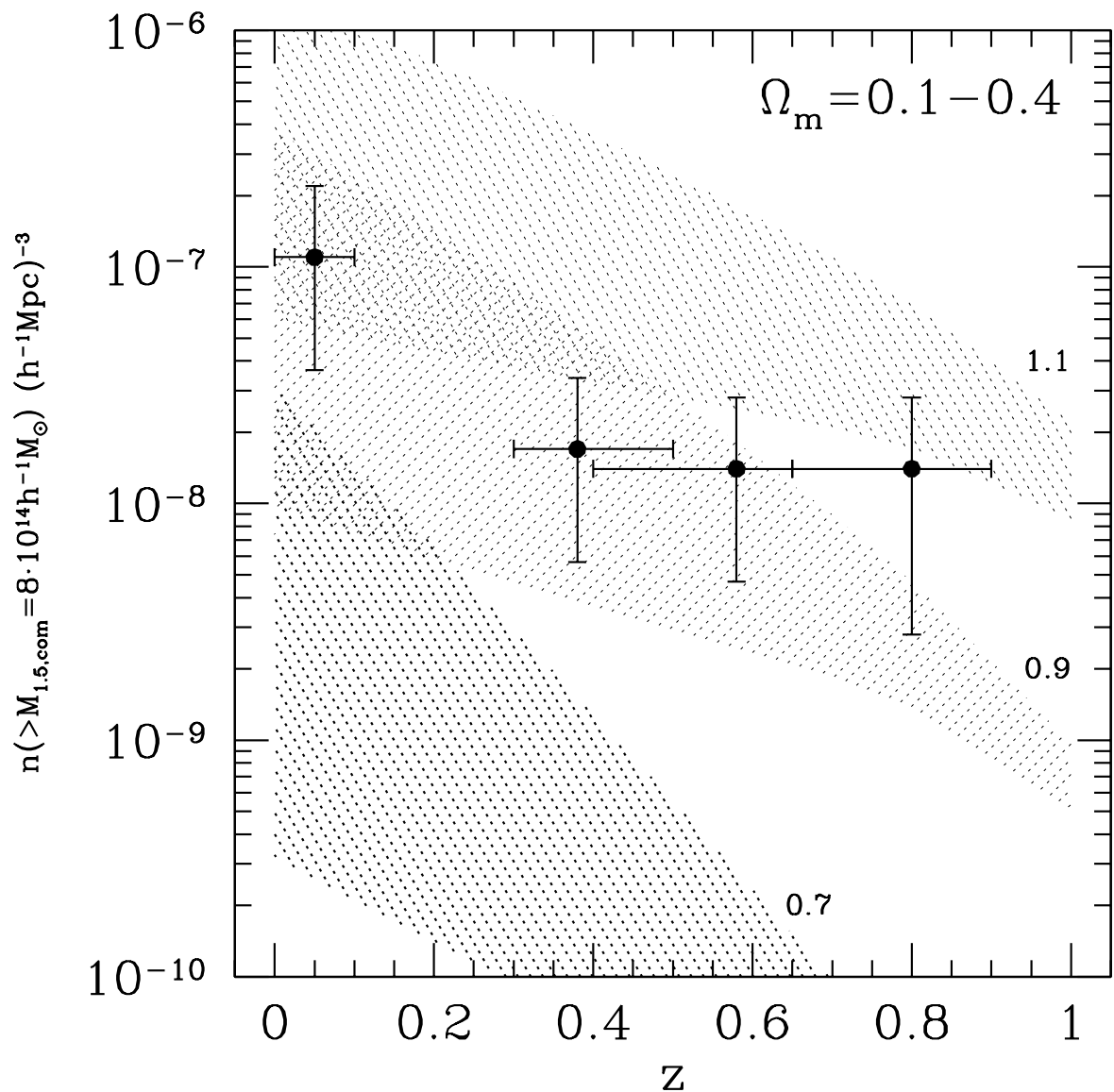}
\caption{Same as Figure \ref{fig:sig8plot},
but for $0.1\leq \Omega_m \leq 0.4$.  Each band is
labeled with the $\sigma_8$ used to predict the number density,
and within each band $\Omega_m$ is varied from 0.1 (bottom
of band) to 0.4 (top); $h=0.72$ and $n=1$.
\label{fig:bomplot}}
\end{figure}

What if $\Omega_m$ is not 0.3? 
How does this affect the allowed range of $\sigma_8$? 
In Figure \ref{fig:bomplot} we compare 
the data to the cluster evolution expected for 
$\sigma_8$ = 1.1, 0.9, and 0.7 (top to bottom bands),
where the width of each band covers all values 
of $\Omega_m$ from 0.1 to 0.4.
It can be seen that the best fit to the data is again
$\sigma_8 \simeq 0.9-1$. 
Low values
of $\sigma_8 \la 0.7$ appear to be excluded by the data, nearly independent of
$\Omega_m$; they provide significantly fewer 
high mass clusters than observed,
especially at $z>0.5$. This result is consistent with the independent
SDSS cluster mass-function at low redshift which yields 
$\sigma_8 = 0.9^{+0.3}_{-0.2}$ and $\Omega_m =0.19^{+0.08}_{-0.07}$
\citep{SDSSmf03}.
The independent result obtained here from the
high redshift clusters 
(Figures \ref{fig:sig8plot}-\ref{fig:bomplot}) 
provides important confirmation that 
$\sigma_8$ is indeed high ($\sim 0.9-1$). 

Finally, we use the observed cluster abundance at high redshifts 
($z \gtrsim 0.5$), independent of the lower 
redshift points, to determine the best
$\sigma_8 - \Omega_m$ normalization relation from high redshift clusters.
This was done by minimizing $\chi^2$, using the error bars given
in \citet{bah98} plus an additional factor of 20\% to account
for uncertainties in the analytic prediction.
The allowed
68\% and 95\% confidence limits are presented by the solid contours in 
Figure \ref{fig:hizont}. As expected, the
observed abundance of these massive clusters at high redshift
depends mostly on  $\sigma_8$ (\S\ref{sec:evol}); 
the dependence on $\Omega_m$ is very weak.
(The dependence on other parameters - Hubble constant and spectral index --
is also weak, as is discussed below). This fact makes the high redshift
cluster abundance method a powerful tool in constraining $\sigma_8$.
The results in Figure \ref{fig:hizont}
show that for any observationally acceptable 
range of $\Omega_m$, from $\sim 0.1$ to $\sim 0.4$, 
the amplitude remains in the range 
of $\sigma_8 \simeq 0.9-1$.
Values of $\sigma_8 <0.8$ are unlikely; 
they have too little power to form  massive systems at $z>0.5$.

We use the $z=0.5-0.8$ clusters to determine 
the best-fit cluster 
normalization relation from the high-redshift objects; we find:
\begin{mathletters} \label{eqn:bestfit}
\begin{eqnarray}
\sigma_8 = 1.03 - 0.3\Omega_m \pm 10\% 
\ \ \ \ \ \ \ \ \ \ \rm{ , or} \\
\sigma_8 \Omega_m^{0.14} = 0.78 \pm 0.08     
\ \ \ \ \ \ \ \ \ \ \ \ \ \ \ \ \
\end{eqnarray}
\end{mathletters}
(where $\sigma_8$ and $\Omega_m$ refer, as before, to their
$z=0$ value).  The linear relation [\ref{eqn:bestfit}a]
is within 3\% of the true
best-fit $\sigma_8$ over the range $\Omega_m=0.1-1.0$; 
the power-law fit [\ref{eqn:bestfit}b]
is superior to equation [\ref{eqn:bestfit}a]
in the range $\Omega_m=0.2-0.7$.
These relations assume a Hubble 
constant of $h=0.72$ and 
spectral index $n=1$, but the results are insensitive to reasonable 
changes in $h$ and $n$. We find that changing 
$h$ by $\Delta h = \pm 0.13$, i.e., from $h=0.59$ to $h=0.85$, changes 
$\sigma_8$ by less than 
$\pm $5\%  (at a given $\Omega_m$); $\sigma_8$ increases slightly 
with $h$. Similarly, changing the spectral index $n=1$ by 
$\Delta n = \pm 0.2$,
from $n=0.8$ to $n=1.2$, 
changes $\sigma_8$ by less than $\pm$5\% (in the same direction: 
slightly higher $\sigma_8$ for higher $n$). 
The 1- and 2-sigma contours of the allowed
$\sigma_8$ - $\Omega_m$ parameter region that includes both these
$h$ and $n$ variations
are presented by the dotted curves in 
Figure \ref{fig:hizont}. 
As expected, these conservative ranges in $h$ and $n$ broaden
the allowed region.
The above results include the estimated uncertainty in the mass threshold 
for the clusters \citep{bah98}. As discussed in \S\ref{sec:evol}, 
only a lower threshold 
(selected at the lower 1-sigma level of the mass estimates)
is used in this analysis, not the individual mass of each cluster.
Reducing the cluster mass threshold further by 10\% 
will reduce the amplitude $\sigma_8$ by $\sim$10\% \citep{FBC97,bah98}.

Also shown in Figure \ref{fig:hizont} are the analogous confidence
contours from the cluster 
mass-function normalization at $z=0.1-0.2$ \citep{SDSSmf03};
the best-fit relation is approximated by
\begin{equation} \label{eqn:sdssfit}
 \sigma_8 \Omega_m^{0.6} = 0.33 \pm 0.03  
\end{equation}
which has a steeper $\Omega_m$ dependence than the  high-redshift
constraint.  These two independent constraints 
overlap only at low $\Omega_m$.  Requiring that both of the
constraints of equations [\ref{eqn:bestfit}] and [\ref{eqn:sdssfit}]
be simultaneously satisfied yields
\begin{mathletters} \label{eqn:bestpar}
\begin{eqnarray}
  \sigma_8 = 0.98 \pm 0.1  \\
  \Omega_m = 0.17 \pm 0.05 
\end{eqnarray}
\end{mathletters}
for the allowed 1-sigma overlap region when $h$=0.72 and $n$=1.
Allowing for variations in $h$ and $n$ as discussed above yields
the same central values for $\sigma_8$ and $\Omega_m$, but 
broadens the allowed range.  Even with the very broad ranges
adopted for $h$ and $n$, the conservative limits  $\sigma_8 > 0.70$
and $\Omega_m < 0.36$ can be set at the 95\% confidence level.

\section{Discussion and Conclusions} \label{sec:disc}

We use the observed abundance of high-mass clusters of galaxies 
at $z=0.5-0.8$ to determine the linear amplitude of mass fluctuations,
$\sigma_8$. The cluster abundance depends exponentially on this amplitude,
and only weakly on other parameters; it therefore provides a powerful method 
for measuring this important parameter.
We show that the relatively high abundance of massive clusters observed
at $z \gtrsim 0.5$, as well as their relatively slow evolution with time, 
requires a high amplitude of mass fluctuations, $\sigma_8 \sim 0.9-1$.
This conclusion is nearly independent of the exact value of $\Omega_m$ 
(in the typical range of $\Omega_m \sim 0.1-0.4$). 

We use the observed abundance at $z\gtrsim 0.5$ to determine a normalization 
relation from high redshift clusters. The relation depends only weakly on
$\Omega_m$: $\sigma_8\Omega_m^{0.14} = 0.78\pm 0.08$; 
alternatively, a linear relation of the form  
$\sigma_8 =1.03-0.3\Omega_m$ ($\pm$10\%)
provides a similarly good fit to the data.
These fits illustrate that 
$\sigma_8 \gtrsim 0.8$ 
for any $\Omega_m \leq 0.4$.  
For the typical observationally suggested value 
of $\Omega_m \simeq 0.2-0.3$, the amplitude is $\sigma_8 =0.95\pm 0.1$.
We emphasize that this high $\sigma_8$ value indicated by the
cluster abundance at high redshift is nearly independent of the exact
value of $\Omega_m$.

\begin{figure}
\plotone{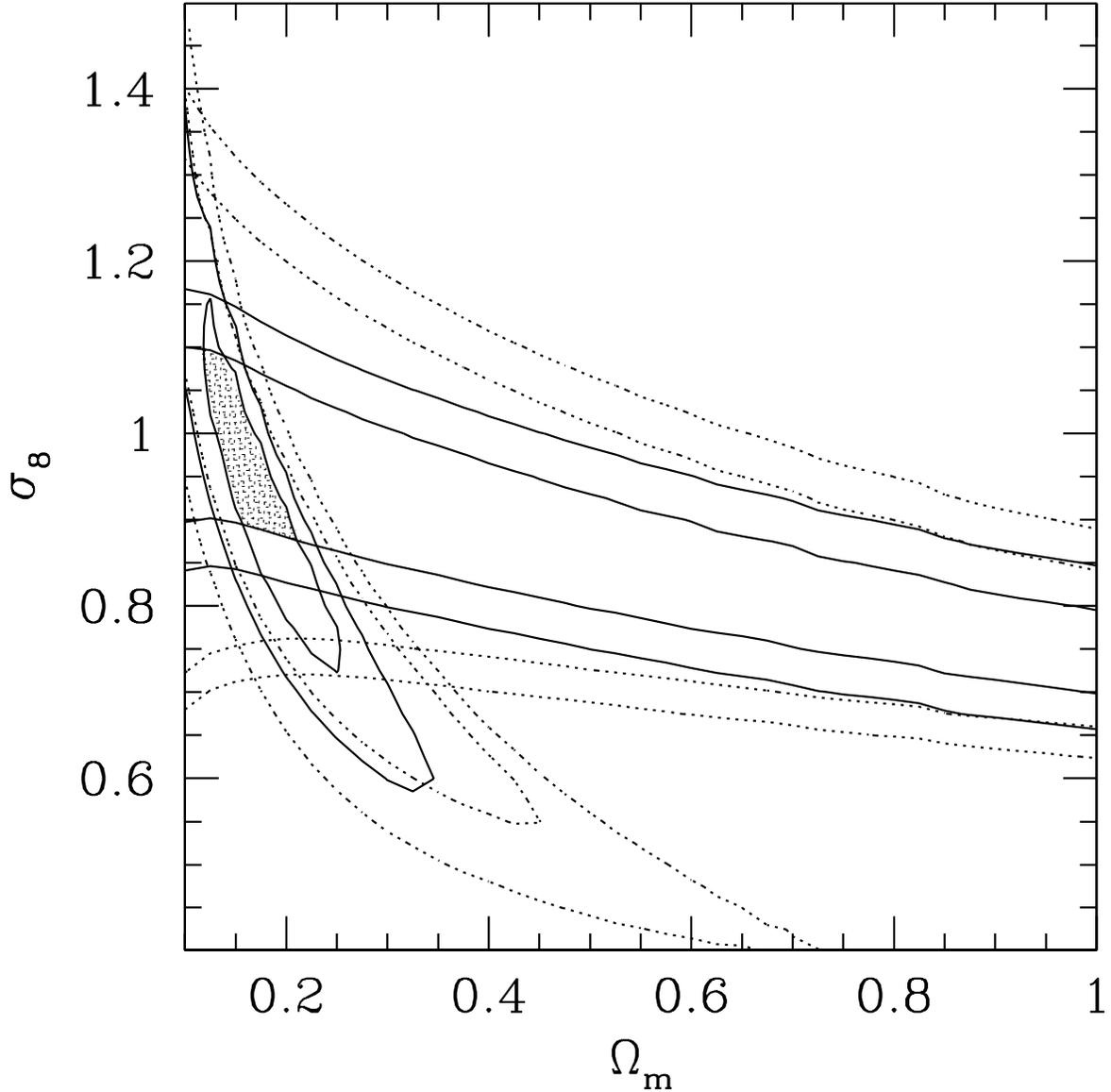}
\caption{Confidence contours in the $\Omega_m-\sigma_8$ plane.
Solid lines are the 68\% and 95\% confidence contours found
from the high-redshift cluster abundance
(contours extending to high $\Omega_m$)
and from the low-redshift SDSS HMF clusters
(extending to lower $\sigma_8$); these assume $h=0.72$ and $n=1$ and
two degrees of freedom.  The shaded region satisfies both constraints
at the 68\% level.
The dotted lines are the 68\% and 95\% limits
with four degrees of freedom
when allowing $0.59<h<0.85$ and $0.8<n<1.2$.
\label{fig:hizont}}
\end{figure}

We combine the high redshift constraint above with the 
independent normalization relation obtained from low redshift cluster 
abundance--- a relation that 
is steeper in $\Omega_m$ ($\sim \Omega_m^{0.6}$; equation [\ref{eqn:sdssfit}]).
The combination breaks the degeneracy between the two parameters. We find 
$\sigma_8 =0.98\pm 0.1$ and $\Omega_m =0.17\pm 0.05$ 
(Figure \ref{fig:hizont}; equation [\ref{eqn:bestpar}]).

The high value of $\sigma_8$ required to explain the high abundance of the
most massive clusters at $z \gtrsim 0.5$ is consistent with 
the present day cluster 
mass function if the mass density parameter is low, $\Omega_m \sim 0.2$. 
If $\Omega_m$=0.3, then the high redshift clusters
still require a high amplitude ($\sigma_8 =0.92 \pm 0.1$), since 
this constraint is nearly independent of $\Omega_m$; the
low redshift cluster abundance is consistent with this value
at the 2-sigma level.

These results improve upon \citet{bah98} by using improvements over 
the standard Press-Schechter formula (which does not accurately
reproduce results from cosmological simulations at high redshift),
allowing for changes in $h$ and $n$,
and by using a more recent cluster normalization relation at low
redshift.  The current results yield slightly lower 
values for the cosmological parameters than the previous work 
(the latter suggested $\sigma_8 =1.2\pm 0.22$ and 
$\Omega_m =0.2^{+0.13}_{-0.07}$ (68\%) when using the most massive clusters)
but are consistent with the new values within the error-bars
\citep{bah98,FBC97}.

The excess CMB fluctuations detected
on small scales by the CBI \citep{mason02}
and the BIMA \citep{dawsea02} experiments implies
(if correctly interpreted as being due to the 
S-Z effect from distant clusters)
that $\sigma_8 =1.04 \pm 0.12$ \citep[95\%;][]{komsel02}.
This is in excellent agreement with our current 
conclusions. We note, however, 
that this high amplitude is inconsistent with the lower value of 
$\sigma_8\approx 0.7$ suggested by current CMB data on large scales 
(which is degenerate with the unknown optical depth).
Future CMB observations should clarify
this current inconsistency.
If massive clusters exist with relatively 
high abundance at high redshifts, as suggested by the data used here 
\citep[as well as by deep X-ray surveys, e.g.][]{RBN02},
then these clusters should indeed 
produce the excess S-Z fluctuations observed by the CMB data.

The relatively high abundance of massive clusters observed at
$z\gtrsim 0.5$ provides one of the strongest arguments for a high
amplitude of mass fluctuations, $\sigma_8\simeq 1$.

{\bf Note added March 7, 2003:} 
The recent CMB anisotropy spectrum released by the WMAP team in 
February 2003 nicely confirms the results presented here.  
The constraint from the CMB alone is $\sigma_8 =0.9\pm 0.1$
\citep{SpergWMAP}, in full agreement with the current
high redshift cluster constraint.

\acknowledgments

This research was supported by the National Computational Science Alliance 
under NSF Cooperative Agreement ASC97-40300, PACI Subaward 766.

\clearpage

\begin{thebibliography}{}

\bibitem[Bahcall \& Fan(1998)]{bah98} Bahcall, N.A. \& Fan, X. 1998,
   \apj, 504, 1

\bibitem[Bahcall, Fan, \& Cen(1997)]{BFC97}Bahcall, N.A.,
     Fan, X. \& Cen, R. 1997, \apj, 485, L53

\bibitem[Bahcall \& Sette(2002)]{bahs02} Bahcall, N.A. \& Sette, A.
    2002, Junior Thesis, Princeton University

\bibitem[Bahcall et al.(1999)]{bah99} Bahcall, N.~A., Ostriker, J.~P.,
     Perlmutter, S. \& Steinhardt, P.  1999, Science, 284, 1481

\bibitem[Bahcall et al.(2000)]{bah00} Bahcall, N.~A., Cen, R., Dav{\'e}, R., 
     Ostriker, J.~P. \& Yu, Q. 2000, \apj, 541, 1

\bibitem[Bahcall et al.(2003)]{SDSSmf03} Bahcall, N.A., et al. 2003,
   \apj, in press (astro-ph/0205490)

\bibitem[Bode et al.(2001)]{bod01} Bode, P., Bahcall, N.A., Ford,
   E.B., Ostriker, J.P.  2001, \apj, 551, 15

\bibitem[Bond et al.(2002)]{bondea02} Bond, J.R., et al. 2002,
   \apj, submitted (astro-ph/0205386)

\bibitem[Carlberg et al.(1997)]{CMYE97} Carlberg, R.G., Morris, S.L.,
      Yee, H.K.C. \& Ellingson, E. 1997, \apj, 479, L19

\bibitem[Carlberg, Yee \& Ellingson(1997)]{CYE97} Carlberg, R.G.,
      Yee, H.K.C. \& Ellingson, E. 1997, ApJ, 478, 462

\bibitem[Carlstrom et al.(2001)]{carea01} Carlstrom, J.E. et al. 2001,
   in Constructing the Universe with Clusters of Galaxies, IAP Conference
   Proceedings, ed. F. Durret \& G. Gerbal, preprint (astro-ph/0103480)

\bibitem[Dawson et al.(2002)]{dawsea02} Dawson, K.S., et al. 
       2002, \apj, in press (astro-ph/0206012)

\bibitem[Donahue \& Voit(1999)]{DV99} Donahue, M. \& Voit, G.M. 1999,
       \apj, 523, L137

\bibitem[Eke, Cole \& Frenk(1996)]{ECF96} Eke, V.R., Cole, S. \&
      Frenk C.S.  1996, MNRAS, 282, 263

\bibitem[Fan, Bahcall \& Cen(1997)]{FBC97}Fan, X., Bahcall,
      N.A. \& Cen, R. 1997, \apj, 490, L123

\bibitem[Fischer \& Tyson(1997)]{fis97} Fischer, P. \& Tyson, J.~A.
       1997, \aj, 1 14, 14

\bibitem[Grego(2001)]{Grego01}Grego, L. 2001 \apj, 552, 2

\bibitem[Hamana et al.(2002)]{hamana02} Hamana, T., et al. 2002,
        \apj, submitted (astro-ph/0210450)

\bibitem[Henry et al.(1992)]{Henryea92}Henry, J.P.,
        Gioia, I.M., Maccacaro, T., Morris, S.L., Stocke, J.T. \& 
	Wolter, A.  1992 \apj, 386, 408

\bibitem[Henry(2000)]{Henry00}Henry, J.P. 2000 \apj, 534, 565

\bibitem[Hjorth, Oukbir \& van Kampen(1998)]{HOvK98} Hjorth, J.,
	Oukbir, J. \& van Kampen, E. 1998, MNRAS, 298, L1

\bibitem[Hu \& Kravtsov(2002)]{HuKrav02} Hu, W. \& Kravtsov, A.V.
   2002, \apj, submitted (astro-ph/0203169)

\bibitem[Ikebe et al.(2002)]{ike02} Ikebe, Y., Reiprich, T.~H.,
   B?hringer, H., Tanaka, Y. \& Kitayama, T. 2002, \aap, 383, 773

\bibitem[Jarvis et al.(2003)]{jarvis03} Jarvis, M., et al. 2003,
   \aj, in press (astro-ph/0210604)

\bibitem[Komatsu \& Seljak(2002)]{komsel02} Komatsu, E. \& Seljak, U.
    2002, \mnras, 336, 1256

\bibitem[Luppino \& Gioia(1995)]{LupGio95} Luppino, G.A. \& Gioia, 
    I.M. 1995, \apjl, 445, L77

\bibitem[Mason et al.(2002)]{mason02} Mason, B.S., et al. 2002,
    \apj, submitted (astro-ph/0205384)

\bibitem[Melchiorri et al.(2003)]{MBBS03} Melchiorri, A., Bode, P., 
    Bahcall, N.A., \& Silk, J. 2003, \apjl, in press (astro-ph/0212276)

\bibitem[Netterfield et al.(2002)]{netterfield} Netterfield, C.B.,
  et al. 2002, \apj, 571, 604 

\bibitem[Oukbir \& Blanchard(1997)]{OB97}Oukbir, J. \& Blanchard,
       A. 1997, \aap, 317, 1

\bibitem[Peebles, Daly \& Juszkiewicz(1989)]{PDJ89}Peebles, P.J.E.,
       Daly, R.A. \& Juszkiewicz, R. 1989, \apj, 347, 563

\bibitem[Reiprich \& Bohringer(2002)]{rei02} Reiprich, T.~H. \&
        Bohringer, H. 2002, \apj, 567, 716

\bibitem[Rosati, Borgani \& Norman(2002)]{RBN02}Rosati, P., Borgani, S.
       \& Norman, C. 2002, \araa, 40, 539

\bibitem[Ruhl et al.(2002)]{ruhlea02} Ruhl, J.E., et al. 2002,
	\apj, submitted (astro-ph/0212229)

\bibitem[Seljak(2001)]{sel01} Seljak, U., MNRAS, submitted (astro-ph/0111362)

\bibitem[Seljak \& Zaldarriaga(1996)]{CMBFAST96} Seljak, U. \&
   Zaldarriaga, M.  1996, \apj, 469, 437

\bibitem[Sievers et al.(2002)]{sievers02} Sievers, J.L., et al.
   2002, \apj, submitted (astro-ph/0205387)

\bibitem[Spergel et al.(2003)]{SpergWMAP} Spergel, D.N., et al.
    2003, ApJ, submitted (astro-ph/0302209)

\bibitem[Wilson, et al.(2001)]{wklc01} Wilson, G., Kaiser, N., Luppino, G.
   \& Cowie, L.L. 2001, \apj, 555, 572
    
\end{thebibliography}
\end{document}